\documentclass{article}
\usepackage{spconf,amsmath,graphicx}
\usepackage{color, url}
\usepackage{booktabs}
\usepackage{enumitem}

\title{Vis2Mus: Exploring Multimodal Representation Mapping for Controllable Music Generation}
 \name{Runbang Zhang$^{1}$, Yixiao Zhang$^{2}$, Kai Shao$^{1}$, Ying Shan$^{3}$, Gus Xia$^{1, 4}$}
\address{$^1$ Music X lab, NYU Shanghai\\
 $^2$ Centre for Digital Music, Queen Mary University of London\\
 $^3$ Tencent Inc.
 $^4$ MBZUAI}
\begin{document}
%
\maketitle
\begin{abstract}
In this study, we explore the representation mapping from the domain of visual arts to the domain of music, with which we can use visual arts as an effective handle to control music generation. Unlike most studies in multimodal representation learning that are purely data-driven, we adopt an \textit{analysis-by-synthesis} approach that combines deep music representation learning with user studies. Such an approach enables us to discover \textit{interpretable} representation mapping without a huge amount of paired data. In particular, we discover that visual-to-music mapping has a nice property similar to \textit{equivariant}. In other words, we can use various image transformations, say, changing brightness, changing contrast, style transfer,  to control the corresponding transformations in the music domain. In addition, we released the Vis2Mus system as a controllable interface for symbolic music generation. \footnote{GitHub repo: \url{https://github.com/ldzhangyx/vis2mus.}}
\end{abstract}
\section{Introduction}\label{sec:introduction}

Multimodal representation learning aims to bridge the heterogeneity gap between different modalities \cite{guo2019deep}. Recently, multimodal representation learning has attracted widespread interest in applications such as cross-modal retrieval \cite{radford2021learning, wang2019camp},  and cross-modal generation \cite{ramesh2021zero, hossain2019comprehensive, ren2020self}. For audio, some deep learning-based work has progressed on tasks such as audio captioning \cite{chen2020audio}, speech synthesis \cite{valle2020flowtron}, and music description \cite{manco2021muscaps, zhang2020butter, zhang2022interpreting}.

In this paper, we follow this research path and aim to explore the representation mapping from image to music. We believe that a better understanding of visual-to-music mapping will not only shed light on the theories of machine learning (especially transfer learning) and cognitive science (especially synesthesia) but also have great practical values on music information retrieval and controllable music generation. For most people, music concepts, such as chords and texture, are quite abstract and it is difficult to use these concepts to directly control music generation. On the contrary, images are more “concrete” and can be used as an intuitive handle to preview and guide music generation once a multimodal mappings is established. 

Due to insufficient available paired image-music data, existing data-driven approaches \cite{tan2020automated, rivas2020pix2pitch} have difficulty in learning a mapping directly from data and generate high-quality music from images. To address this problem, we resort to \textit{analysis-by-synthesis} methodology \cite{abs} and develop a method that combines representation learning with user subjective evaluations. In specific, we i) propose several transformations that can be applied to both music and image representations, ii) synthesize image-music pairs data using deep generative models by applying the proposed transformations, and iii) conduct user studies to evaluate the synthesized pairs to explore properties of visual-to-image mappings. 

We find that visual-to-music mapping has a nice property analogous to equivariant. This property enables us to use various image transformations (including changing brightness, changing contrast and style transfer) to control the corresponding transformations in the music domain. In addition, we release the Vis2Mus system, which applies our approach to a large amount of paired image and music as a controllable interface for symbolic music generation.

\section{Methodology} \label{sec: vis2mus}

Given an image, the problem of “what is the ideal corresponding music” may seem too difficult or intractable, and different people probably have different answers. In this paper, we aim to solve a \textit{conditioned} simplification of the problem which is more accessible, in which we: 1) first assume an image-music pair, 2) then modify the image by changing some features, and 3) finally ask “how we should change the music accordingly to match the modified image.” 

Such problem setup is also graphically represented in Figure 1, in which we can regard the given (conditioned) image-music pair as an “anchor” to control or judge any modification or variation based on it. Formally, let $v$ denote the visual input, $m$ denote the music output, and function $f$ be the conceptual visual-to-music mapping. Our problem setting is that given a specific pair $m = f(v)$ and a certain transformation $g$ such that $v'=g(v)$, what is the corresponding music $m'$, i.e.,  $f(v')$?

\begin{figure}[ht]
    \centering
    \includegraphics[width=0.5\linewidth]{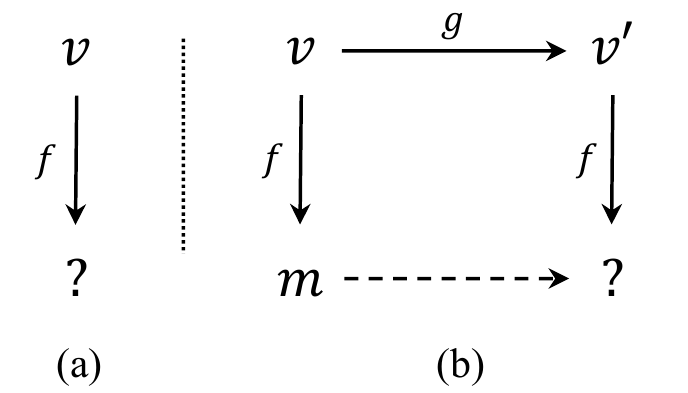}
    \caption{An illustration of problem setup: (a) unconditioned multimodal mapping Vs. (b) conditional multimodal mapping. Our study focuses on the scenario of (b).}
    \label{fig:diagram2}
\end{figure}

We develop an analysis-by-synthesis method that combines deep music generation with user studies. Inspired by Kandinsky's statements on synesthesia\cite{kandinsky2012concerning}, our hypothesis is that $f$ has an isomorphic property with respect to a certain transformations that can be applied in both visual and music domains. Formally, if $m=f(v) $ and $v'=g (v)$, then $m'=g(m) $. In other words, given a pair $(v, m)$, we can infer $m'$ without knowing the exact form of $f$ .

Under this hypothesis, we first propose several concrete forms of $g$. Then, we synthesize the corresponding $m'$ via deep music generation. Finally, we conduct user studies to see whether users can distinguish $m'$ from other candidates. In the rest of this section, we introduce the deep generative model and how to apply $g$ on its latent representation in section 2.1, discuss several particular forms of $g$ in section 2.2 and section 3.3, followed which we present the user study and experimental results in section 2.4 and section 2.5. 

\subsection{The Backend Deep-generative Model}

\begin{figure}[ht]
    \centering
    \includegraphics[width=\linewidth]{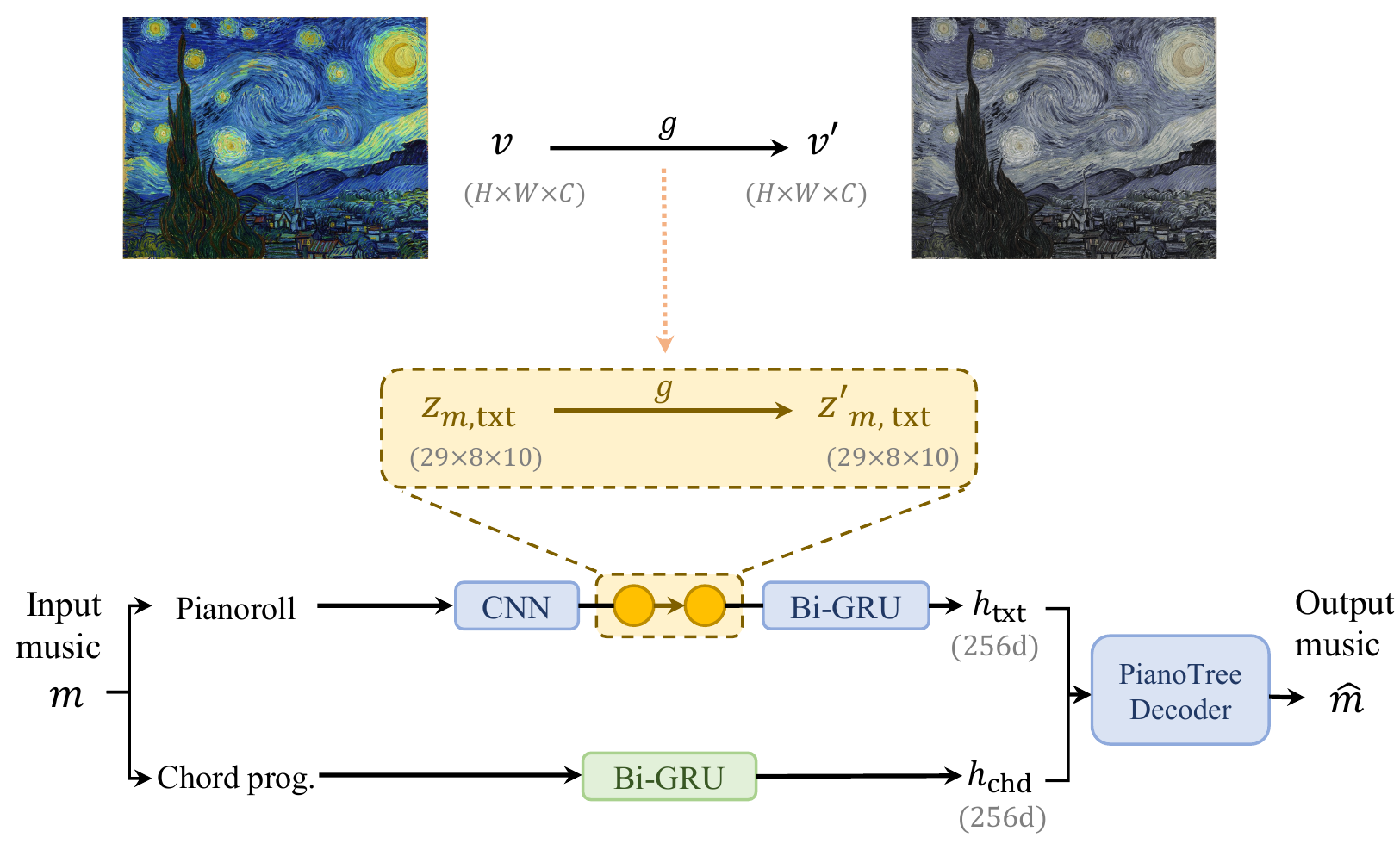}
    \caption{An illustration of the polyphonic disentanglement model and how to apply cross-modal transformation $g$ to the latent texture feature.}
    \label{fig:model}
\end{figure}

In order to generate music from the latent space, we use a pretrained polyphonic music representation learning model (the bottom half of Figure \ref{fig:model}) \cite{wang2020learning}, which can learn disentangled chord and texture representations from music, as our backend model. Note that “texture”(as in English) can refer to both visual and musical features, and our hypothetical cross-modal transformation $g$ acts on the latent texture feature. 

The music model uses two separate encoders, a texture encoder and a chord encoder, to learn corresponding latent representations, $h_\text{txt}$ and $h_\text{chd}$. In specific, the texture encoder takes a quasi-piano-roll input and consists a CNN layer followed and a bi-directional GRU. The chord encoder takes a chord progression input and is implemented by a bi-directional GRU. The model also uses a PianoTree decoder \cite{wang2020pianotree} to reconstruct the music from the two representations.

The upper half of Figure \ref{fig:model} shows an example of image transformation. We let the same transformation $g$ acts on both image $v$ and the the feature map $z_{m, \text{txt}}$, the intermediate output of the CNN module of the texture encoder. The rationale of such design is that the intermediate representation of the music texture $z_{\text{txt}, m}$ is a two-dimensional feature map, which makes it easy to perform various image-like operations on it, such as changing the "brightness" and "contrast" of the feature map. Finally, The manipulated feature map $z'_{\text{txt}, m}$ is sent back to the music model to reconstruct a new piece of music with the transferred representation.


\subsection{Brightness and Contrast Transformation}

\subsubsection{Brightness Transformation}

The brightness transformation $g_b$ can be regarded as a pixel-wise function. Let $x$ be any  channel (a matrix) of the $z_m$ or $v$, then for every element (the pixel value) of $x_i$, its brightness transformation is defined as: 
\begin{equation}
x_i' = g_b(x_i) = x_i+ \alpha_b/2\cdot\max(x) 
\end{equation}
where $\alpha_b$ is a hyperparameter. Figure 4 shows an example of synthesizing a new image-music pair $(v', m')$ based on an existing pair $(v, m)$ using brightness transformation.

\begin{figure}[ht]
    \centering
    \includegraphics[width=0.5\linewidth]{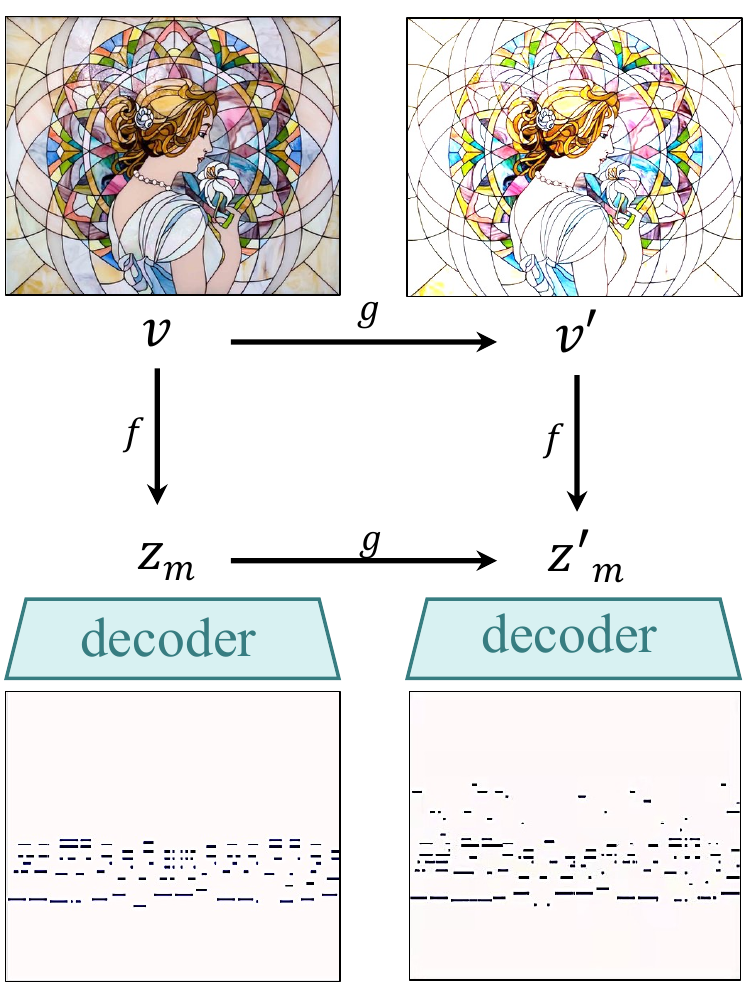}
    \caption{An illustration of applying cross-modal brightness transformation $g_b$ on image $v$ and music $m$, respectively, in which $(v, m)$ is a matched image-music pair.}
    \label{fig:brightness}
\end{figure}


\subsubsection{Contrast Transformation}

Similarly to brightness transformation, the contrast  transformation $g_c$ is defined as:
\begin{equation}
 x_i' = g_c(x_i) =  \overline{x} +(x_i- \overline{x})\cdot(1+\alpha_c)   
\end{equation}
where $\alpha_c$ is a hyperparameter. Figure 5 shows an example of synthesizing a new image-music pair $(v', m')$ based on an existing pair $(v, m)$ using contrast transformation.

\begin{figure}[ht]
    \centering
    \includegraphics[width=0.5\linewidth]{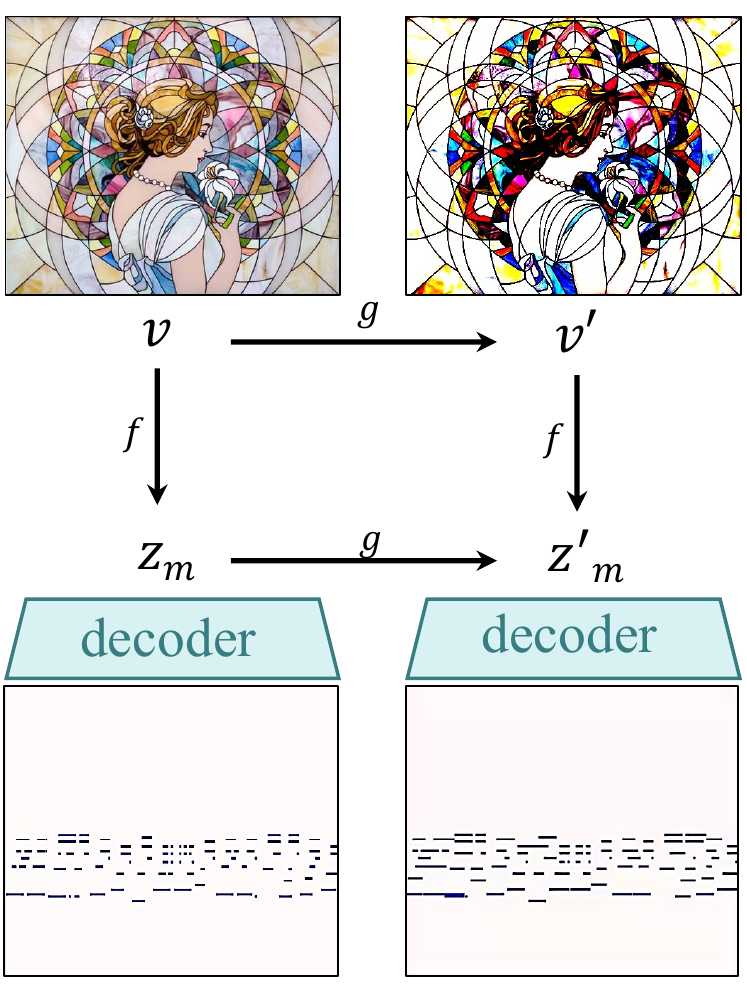}
    \caption{An illustration of applying cross-modal contrast transformation $g_c$ on image $v$ and music $m$, respectively, in which $(v, m)$ is a matched image-music pair.}
    \label{fig:contrast}
\end{figure}


\subsection{Style Transfer}

Besides changing brightness and contrast, we also consider style transfer as a special form of transformation. In general, style transfer is achieved by fusing a content element and a style element. In the image domain, people usually regard image contour as content and a colored texture as style \cite{jing2019neural}. Correspondingly, we regard melody contour as content and polyphonic texture as style in the music domain. 

Formally, let $z=(z_c, z_\text{txt})$ be the representation of an image or a piece of music, where $z_c$ and $z_\text{txt}$ represent the corresponding contour and texture, respectively. The cross-modal style transfer operation $g_s$generates the latent representation of a new image or music:
\begin{equation}
z' = g_s(z) = g_s(z_c, z_\text{txt})
\end{equation} where $g_s$ simply concatenate contour and texture representation.

\begin{figure}[ht]
    \centering\includegraphics[width=0.85\linewidth]{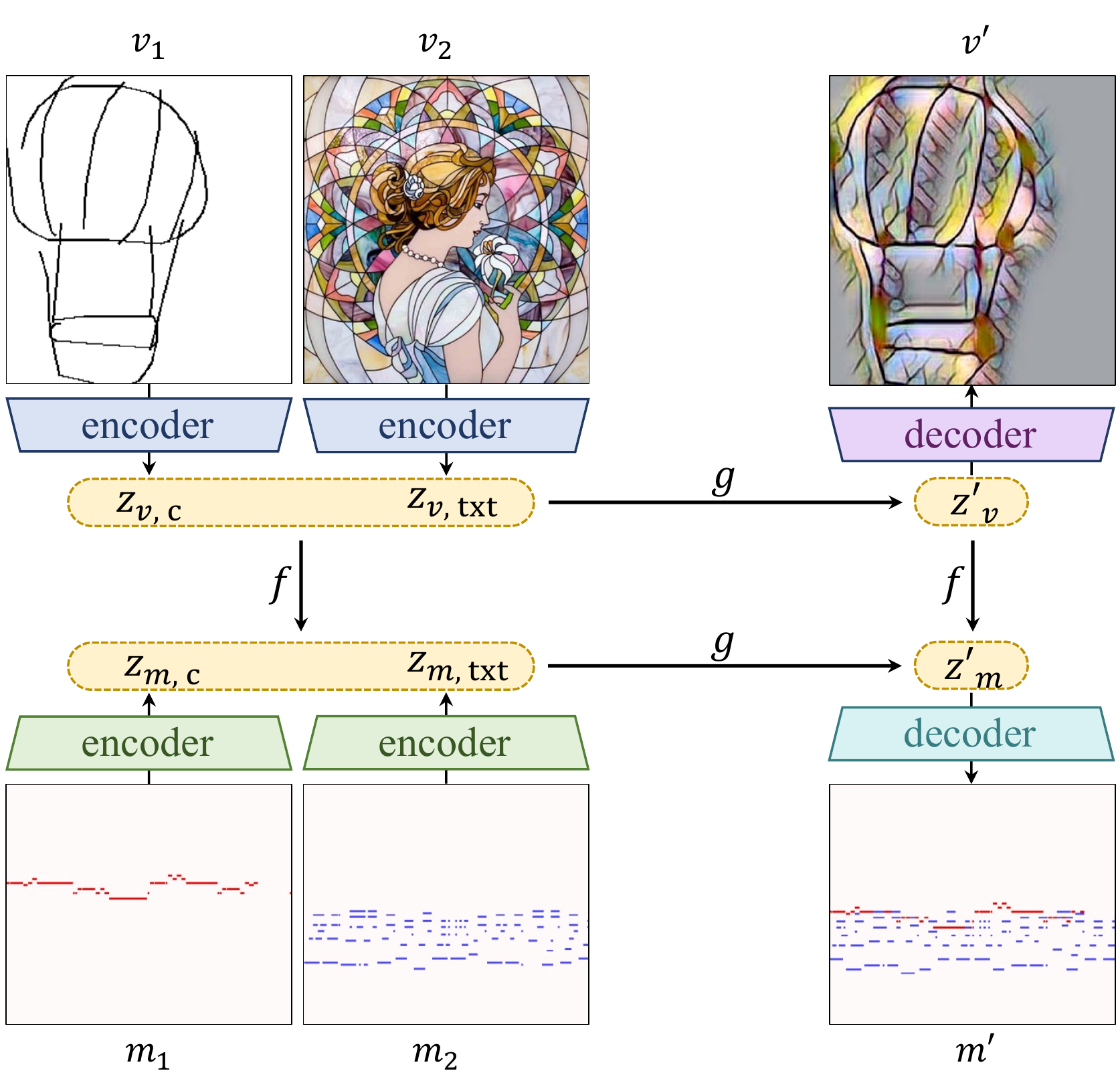}
    \caption{An illustration of applying cross-modal style transformation $g$ on the images $(v_1, v_2)$ and the music pieces $(m_1, m_2)$, respectively, in which $(v_1, m_1)$ and  $(v_2, m_2)$ are two matched image-music pairs.}
    \label{fig:transfer}
\end{figure}

Figure 5 shows an example of synthesizing a new image-music pair $(v', m')$ based on $(v_1, v_2)$ and $(m_1, m_2)$. Here, $v_1$ is a sketched contour, $v_2$ is a styled image (with colored texture), $m_1$ is a melody contour, and $m_2$ is a styled accompaniment (with polyphonic texture). The encoder and decoder of the image style transfer part come from \cite{zhang2017multistyle}, while the encoder and decoder of the music style transfer part come from \cite{zhao2021accomontage, wang2020learning}. In specific, the compositional style transfer algorithm first adjusts the pitch register and the tempo of the melody and accompaniment, respectively, to let them match each other. Then, our backend deep-generative model (as described in section 2.1) performs chord style transfer \cite{wang2020learning} to the accompaniment by substituting the latent representation of original chords with the latent representation of melody’s chords. In this way, the music texture is fused into the melody, just as the style of the image is fused into the sketch.

\subsection{User Study Design}

The user study contains three parts: 1) brightness transformation user study, 2) contrast transformation user study, and 3) style transfer user study. The brightness transformation user study and the contrast transformation user study have the same three steps:

\begin{itemize}[parsep=0pt,  topsep=0pt]
    \item[1.] \textbf{Preview}: The user is asked to view and listen to an image-music pair; 
    \item[2.] \textbf{Transformation}: The image is transformed in terms of brightness or contrast, while the music is altered both by the same transformation and the inverse transformation, producing two new pieces of music. E.g. if the image is transformed to be brighter, the inverse transformation means to make the latent representation of the paired music dimmer;
    \item[3.] \textbf{Test}: The user is asked to select which of the two new pieces better matches the transformed image. Therefore, the corresponding random guess baseline accuracy is 50\%.
\end{itemize}

The style transfer user study also has 3 steps:

\begin{itemize}[parsep=0pt,  topsep=0pt]
    \item[1.] \textbf{Preview}:The user is asked to view and listen to four image-music pairs: two sketch-melody pairs and two image-accompaniment pairs.
    \item[2.] \textbf{Transformation}: A synthesized image is shown, which is transformed using a randomly selected one out of the two sketches as the content and a randomly selected one of the two images as the style. At the same time, four pieces of synthetic music are displayed, generated by all possible combinations of the two melodies and the two accompaniments shown in the first step using compositional style transfer.
    \item[3.] \textbf{Test}: The user is asked to select one synthetic music from the four that perceptually best matches the synthesized image. Therefore, the corresponding random guess baseline accuracy is 25\%.
\end{itemize}

\subsection{Experimental Results}

For the brightness user study and the contrast user study, the gender distribution of all subjects was 61\% male and 39\% female, and the music level distribution was 36\% amateur, 39\% intermediate, and 25\% professional; a total of 24 and 22 responses were collected respectively. For the style transfer user study, the gender distribution was 76\% male and 24\% female, and the music level distribution was 39\% amateur, 46\% intermediate, and 15\% professional; a total of 61 responses were collected. Table \ref{tab:exp1} shows the results of three user studies, with user selection accuracy significantly higher than baseline random guesses in all tests. 

\begin{table}[ht]
\small
    \centering
\begin{tabular}{lccc}
\toprule
  Acc (\%)       & \small{Brightness} & \small{Contrast} & \small{Style} \\
\midrule
Random   & 50.0       & 50.0     & 25.0         \\
\midrule
 Ours & \textbf{62.5}$^\ast$       & \textbf{68.2}$^\ast$      &  \textbf{60.6}$^\ast$       \\
\bottomrule
\end{tabular}
    \caption{Evaluation results of the transformation experiments. $\ast$: The accuracy outperforms random guesses with $p < 0.05$ on \textit{binomial test}.}
    \label{tab:exp1}
\end{table}

This result indicates that given an image-music pair $(m,v)$ and a transformed image $v'$ with respect to brightness, contrast or style, humans can tell the corresponding music $m'$ that is ``brighter''/``dimmer'', with more/less ``contrast'', or style transformed \textit{without} any training. The commutative cross-modal transformations do exist, though not for everyone, yet statistically significant. Such property also enables us to use image transformations to control corresponding music transformations.

\section{Vis2Mus: The User Interface} \label{sec:system}

\begin{figure}[ht]
    \centering
    \includegraphics[width=\linewidth]{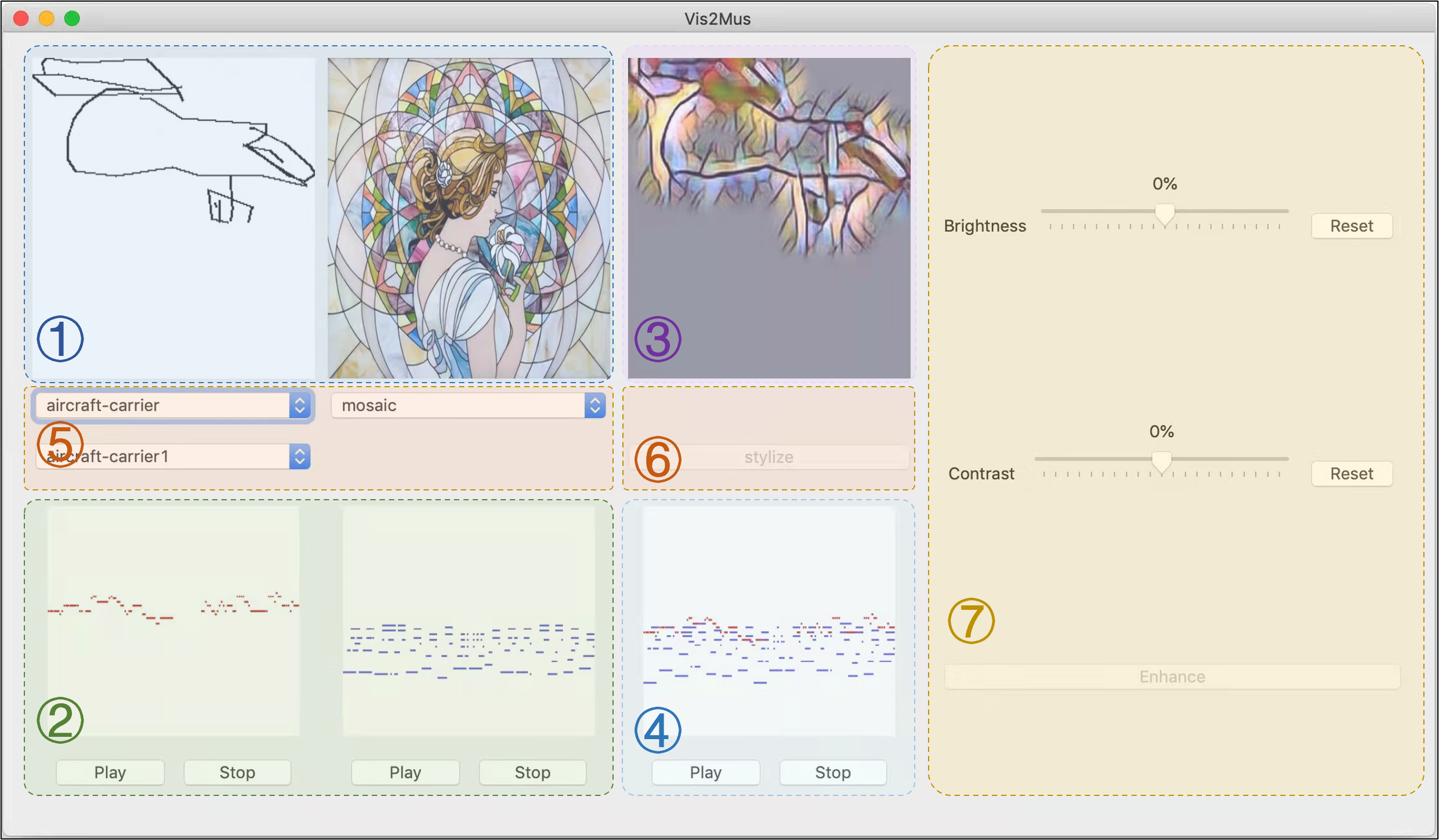}
    \caption{A display of the interface of the Vis2Mus system. }
    \label{fig:system}
\end{figure}

Based on the visual-to-music mapping, we develop the Vis2Mus system as a controllable music generation interface, as shown in Figure \ref{fig:system}. Users can generate music by following the steps below:

\begin{itemize}[parsep=0pt, itemsep=0pt, topsep=0pt]
    \item[1.] \textbf{Select initial melody-sketch and polyphony-image} anchor pairs from drop-down boxes 
 in \textcircled{\raisebox{-0.9pt}{5}}. The selected items (which are prelabeled) will be displayed in \textcircled{\raisebox{-0.9pt}{1}} and  \textcircled{\raisebox{-0.9pt}{2}};
    \item[2.] \textbf{Conduct style transfer} as in Section 2.3. Pressing button \textcircled{\raisebox{-0.9pt}{6}} and the system will generate a fused image (displayed in \textcircled{\raisebox{-0.9pt}{3}}) as a visualization and preview of the fused music piece (displayed in \textcircled{\raisebox{-0.9pt}{4}}).
    \item[3.] \textbf{(Optional) Tuning brightness and contrast of generated music} as in Section 2.1 and 2.2. Changing the brightness and contrast of the style-transferred image in \textcircled{\raisebox{-0.9pt}{7}} to further control the generated music in \textcircled{\raisebox{-0.9pt}{4}}.
\end{itemize}


\vspace{-0.4cm}

\section{Conclusion and Future Work}

In this study, we have contributed an analysis-by-synthesis method, which combines deep generative modeling and user study to explore interpretable visual-to-music mapping in a data efficient way. We discovered that visual-to-music mapping in the latent space has a nice property analogous to equivariant with respect to three transformations: changing brightness, changing contrast, and style transfer. The followup user study shows that the results are significant higher than random guess baselines. Inspired by the discoveries, we have also developed Vis2Mus, a controllable music generation interface using images and image transformations as the control handlers. 

We are well-awared that we have just explored a small portion of visual-to-music mapping. In the future, we plan to study more forms of cross-modal transformation acting on more possible latent factors. We also plan to conduct a much larger scale user study, from which more interesting multimodal patterns can potentially be discovered. 

\bibliographystyle{IEEEbib}
\bibliography{strings,refs}

\end{document}